\documentclass[twocolumn,prl,showpacs,preprintnumbers,amsmath,amssymb,superscriptaddress]{revtex4}
\usepackage{graphicx}
\usepackage{setspace}

\usepackage{color}
\usepackage{ulem} 

\usepackage{amsmath,amsthm,amssymb}
\usepackage{mathrsfs}

\newcommand{\sigx}{\hat{\sigma}_1}

\newcommand{\sigz}{\hat{\sigma}_3}

\makeatletter
\def\Left#1#2\Right{\begingroup%
   \def\ts@r{\nulldelimiterspace=0pt \mathsurround=0pt}%
   \let\@hat=#1%
   \def\sht@im{#2}%
   \def\@t{{\mathchoice{\def\@fen{\displaystyle}\k@fel}%
          {\def\@fen{\textstyle}\k@fel}%
          {\def\@fen{\scriptstyle}\k@fel}%
          {\def\@fen{\scriptscriptstyle}\k@fel}}}%
   \def\g@rin{\ts@r\left\@hat\vphantom{\sht@im}\right.}%
   \def\k@fel{\setbox0=\hbox{$\@fen\g@rin$}\hbox{%
      $\@fen \kern.3875\wd0 \copy0 \kern-.3875\wd0%
      \llap{\copy0}\kern.3875\wd0$}}%
      \def\pt@h{\mathopen\@t}\pt@h\sht@im%
      \Right}%
\def\Right#1{\let\@hat=#1%
   \def\st@m{\mathclose\@t}%
   \st@m\endgroup}
\makeatother


\begin{document}
\title{Photo-induced enhancement of excitonic order}

\author{Yuta Murakami}
\affiliation{Department of Physics, University of Fribourg, 1700 Fribourg, Switzerland}
\author{Denis Gole\v z}
\affiliation{Department of Physics, University of Fribourg, 1700 Fribourg, Switzerland}
\author{Martin Eckstein}
\affiliation{Department of Physics, University of Erlangen-N\"urnberg, 91058 Erlangen, Germany}
\author{Philipp Werner}
\affiliation{Department of Physics, University of Fribourg, 1700 Fribourg, Switzerland}
\date{\today}

\begin{abstract}

We study the dynamics of excitonic insulators coupled to phonons.
Without phonon couplings, the linear response is given by the damped amplitude oscillations of the order parameter with frequency equal to the minimum band gap. 
A phonon coupling to the interband transfer integral induces two types of long-lived collective oscillations of the amplitude, one originating from the phonon dynamics and the other from the phase mode, which 
becomes massive. 
We show that even for small phonon coupling, a photo-induced enhancement of the exciton condensation and the gap can be realized.  
Using the Anderson pseudo-spin picture, we argue that the origin of the enhancement is a cooperative effect of the massive phase mode and the Hartree shift induced by the photo excitation.
We also discuss how the enhancement of the order and the collective modes can be observed with time-resolved photo-emission spectroscopy.
\end{abstract}

\maketitle
{\it Introduction--}
Nonequilibrium dynamics can provide new insights into properties of materials, and new ways to control ordered states.
In this context, superconducting (SC) phases have been studied extensively.
Both a light-induced enhancement of SC \cite{fausti2011,kaiser2014,mitrano2016} and the observation of the amplitude Higgs mode \cite{Matsunaga2013,Matsunaga2014,Matsunaga2017} have been reported. 
Recently, also a related family of ordered states, i.e. excitonic insulators (EIs), has attracted interest
\cite{Rohwer2011,Hellmann2012,Porer2014,Denis2016,Mor2016,Kaiser2016}.
An EI state is formed by the macroscopic condensation of electron-hole pairs \cite{jerome1967,kohn1967}, and its theoretical description is analogous to the BCS or BEC theory for SC.
Although the pioneering idea of 
exciton condensation 
was proposed in the 1960s \cite{kohn1967,jerome1967,halperin1968},
the interest in this topic has been renewed recently by the study of some candidate materials such as $1T$-TiSe$_2$ and Ta$_2$NiSe$_5$ (TNS) \cite{Monney2007,monney2009,Wakisaka2009,Wakisaka2012,Ohta2014,Lu2017}. 
Their analogy to SC makes the EI an interesting system for nonequilibrium studies.
In particular, for TNS, time and angle resolved photo-emission spectroscopy (trARPES) experiments showed that the direct band gap can be either decreased or increased depending on the pump fluence \cite{Mor2016}, which was interpreted in terms of a photo-induced enhancement or suppression of excitonic order. 
A more recent report of the amplitude mode \cite{Kaiser2016} provides further confirmation of an EI state in TNS.

So far the theoretical works on EIs have mainly focused on the equilibrium properties such as the BEC-BCS crossover \cite{phan2010,zenker2012}, the coupling of the EI to phonons \cite{kaneko2013,zenker2014,monney2011}, linear susceptibilities \cite{Sugimoto2016,Sugimoto2016b,Matsuura2016}, and the effect of strong interactions and new emergent phases \cite{balents2000,kunes2015,Nasu2016}. In contrast, the nonequilibrium investigation of EIs has just begun \cite{Denis2016}. 
In this work, using TNS as a model system, we clarify two basic effects of the electron-phonon (el-ph) coupling on the dynamics of EIs,
(i) the effect on collective modes, and (ii) the impact on the excitonic order after photo-excitation.
In particular, we show that with phonons photo-excitation can result in an enhancement of the order. 

{\it Formalism--}
In this paper we focus on a two-band model of spin-less electrons coupled to phonons, 
\begin{align}
\hat{H}(t)&=\sum_{{k},\alpha=0,1} (\epsilon_{k,\alpha}+\Delta_{\alpha})\hat{c}^\dagger_{{k},\alpha}\hat{c}_{{k},\alpha}\nonumber\\
&+E(t)\sum_{k}(\hat{c}^\dagger_{{k},1}\hat{c}_{{k},0}+\hat{c}^\dagger_{{k},0}\hat{c}_{{k},1})+U\sum_i \hat{n}_{i,0}\hat{n}_{i,1}\\
&+g\sum_i (\hat{b}_{i}^\dagger+\hat{b}_{i})(\hat{c}^\dagger_{i,1}\hat{c}_{i,0}+\hat{c}^\dagger_{i,0}\hat{c}_{i,1})+\omega_0\sum_i \hat{b}_{i}^\dagger \hat{b}_{i}\nonumber.
\end{align}
In order to mimic the quasi-one dimensionality and direct band gap in TNS, we consider a one-dimensional configuration with $N$ sites ($N\rightarrow \infty$).
$\hat{c}^\dagger_{i, \alpha}$ is the creation operator of an electron at site $i$ in band $\alpha$ $(=0,1)$, 
$k$ indicates the momentum and $c_{k,\alpha}^\dagger=\frac{1}{\sqrt{N}}\sum_j e^{ik\cdot j} c_{j,\alpha}^\dagger$.
$0$ and $1$ indicate the conduction and valence bands respectively, and $\hat{n}_{i,\alpha}=\hat{c}^\dagger_{i,\alpha}\hat{c}_{i,\alpha}$. 
$\Delta_{\alpha}$ is the crystal field, and we choose $\epsilon_{k,\alpha}=-2J(-1)^\alpha \cos(k)$. 
$E(t)$ denotes the external laser field, and we assume that the laser couples to the system through dipolar transitions.
$U$ is the interaction between bands, and the driving force for the excitonic pairing.
We denote the phonon creation operator at site $i$ by $\hat{b}_{i}^\dagger$, the el-ph coupling by $g$, the phonon frequency by $\omega_0$, and define the effective el-ph interaction by $\lambda\equiv2g^2/\omega_0$.  
This phonon is associated with a lattice distortion in TNS, which hybridizes the electronic bands \cite{kaneko2013}. 
The hopping parameter $J$ is our unit of energy. 

For $g=0$, the system has a $U(1)$ symmetry (corresponding to the conservation of the particle number in each band), which is broken in the EI.  In the el-ph coupled system ($g\ne 0$), this $U(1)$ symmetry is reduced to a $Z_2$ symmetry,
and we will see that this has profound consequences for the collective modes and the nonequilibrium dynamics. 

The dynamics of the system is studied within time-dependent mean-field theory at $T=0$~\cite{supplemental}.
We define the order parameter of the excitonic condensate as $\phi(t)\equiv \langle \hat{c}_{i,0}^\dagger(t) \hat{c}_{i,1}(t)\rangle$, which we take real in equilibrium,
the difference in occupancy between the conduction and valence bands $\Delta n(t)\equiv\langle \hat{n}_{i,0}(t)\rangle -\langle \hat{n}_{i,1}(t)\rangle$, and the phonon displacements $X(t)\equiv \langle \hat{b}_{i}^\dagger(t)+\hat{b}_{i}(t)\rangle$. 
The choice of $i$ does not matter because of the homogeneous excitation.
The mean-field time evolution is self-consistently determined through $\phi(t)$, $\Delta n(t)$ and $X(t)$.
This can be simply expressed in a pseudo-spin representation,
$\hat{S}^\gamma_{{k}}\equiv
\hat{\Psi}^\dagger_{{k}}
\frac{1}{2}\hat{\sigma}_\gamma 
\hat{\Psi}_{{k}}$ with the spinor $\hat{\Psi}^\dagger_k \equiv [\hat{c}^\dagger_{k,0},\hat{c}^\dagger_{k,1}]$, in analogy to the Anderson pseudo-spin representation \cite{Anderson1958}. Here $\hat{\sigma}_{\gamma}$ for $\gamma=x,y,z$ denotes the Pauli matrix and $\gamma=0$ is the identity matrix.
The spin commutation relation is fulfilled, $[\hat{S}^{\alpha},\hat{S}^{\beta}]=i\epsilon_{\alpha,\beta,\gamma}\hat{S}^\gamma$, except for $\hat{S}^0$,  which commutes with all other operators. 
With these operators, $\Delta n(t)=\frac{2}{N} \sum_{k}\langle \hat{S}^z_{k}(t)\rangle$, the total number of particles per site is $\langle \hat{n}_0(t)\rangle+\langle \hat{n}_1(t)\rangle=\frac{2}{N} \sum_{k}\langle \hat{S}^0_{k}(t)\rangle $, and the order parameter is $\phi(t)=\frac{1}{N}\sum_{k}\langle \hat{S}^x_{k}(t)+i\hat{S}^y_{k}(t)\rangle$.

In mean-field, the time evolution of electrons is expressed using the pseudo-spin expectation values as  
\begin{align}
\partial_t {\bf S}_{k}(t)={\bf B}_{k}(t)\times {\bf S}_{k}(t),\label{eq:pseudo_spin_dynamics}
\end{align}
with a pseudo-magnetic field ${\bf B}_{k}(t)$
\begin{subequations}
\begin{align}
B_{k}^x(t)&=2(E(t)+gX(t)-U{\rm Re} \phi(t)),\label{eq:EqBx}\\
B_{k}^y(t)&=-2U{\rm Im} \phi(t),\label{eq:EqBy}\\
B_{k}^z(t)&=2\epsilon_{k}^z-U\Delta n(t),\label{eq:EqBz}
\end{align}
\label{eq:EqB}
\end{subequations}
where $2\epsilon_{k}^z\equiv (\epsilon_{{k},0}-\epsilon_{{k},1})+(\Delta_0-\Delta_1)$.
As for the phonons, 
$\partial_t P(t)=-\omega_0X(t)-2g(\phi(t)+\phi(t)^*)$
with $\partial_t X(t)=\omega_0 P(t)$.
Eqs.~(\ref{eq:EqB}) imply that in the normal state (with $E(t)=0$) the pseudo-magnetic field and pseudo-spins are along the $z$ axis, while in the EI state, there are additional $x$ or $y$ components.
In particular, the phonon contributes to the $x$ component.

{\it Results--}
We choose the reference parameters such that the model with $\lambda=0$ reproduces the ARPES spectra of TNS \cite{Mor2016}, namely $\Delta_0=-0.55$, $\Delta_1=-2.45$, $\lambda=0$ and $U=3$, which is half-filled and in the BEC regime \cite{phan2010}
\footnote{In Ref.~\onlinecite{Ohta2014} another set of parameters is considered, which turns out to be in the BCS regime at $T=0$. 
Even there the light-enhancement of the EI phase can be observed, see the supplemental material.}.
The minimum gap between the quasiparticle bands $\Delta_{\rm EI}$ is located at $k=0$ in equilibrium.
In the calculations with $\lambda>0$ 
we adjust $\Delta_\alpha$, and  $U$ such that in equilibrium the electronic properties (i.e. $\phi$, $\Delta n$ and the quasi-particle dispersion) are the same as for $\lambda=0$. 
In the following we fix $\omega_0=0.1$.

We start with the linear response regime and investigate the collective modes at $q=0$ by analyzing the linear susceptibility
$\chi_{11}^R(t)\equiv-i\theta(t)N\langle [\hat{\rho}_{1}(t),\hat{\rho}_{1}(0)]\rangle$,
with $\hat{\rho}_{1}=\frac{1}{N}\sum_i (\hat{c}^\dagger_{i,1}\hat{c}_{i,0}+\hat{c}^\dagger_{i,0}\hat{c}_{i,1})$ 
\footnote{We evaluate the susceptibility by following  $2\text{Re}\phi(t)$ after an initial perturbation 
$E(t)=d_f\delta(t)$ with small enough $d_f$.}.
Since the initial $\phi$ is real, $\chi_{11}^R$ captures the dynamics of the amplitude of the order parameter $\phi(t)$. 
Note that $\chi_{11}^R$ corresponds to the dynamical pair susceptibility in SCs, which has been used to study the amplitude Higgs mode~\cite{Kulik1981,Murakami2016,Murakami2016b}.

 \begin{figure}[t]
  \centering
    \hspace{-0.cm}
    \vspace{0.0cm}
   \includegraphics[width=85mm]{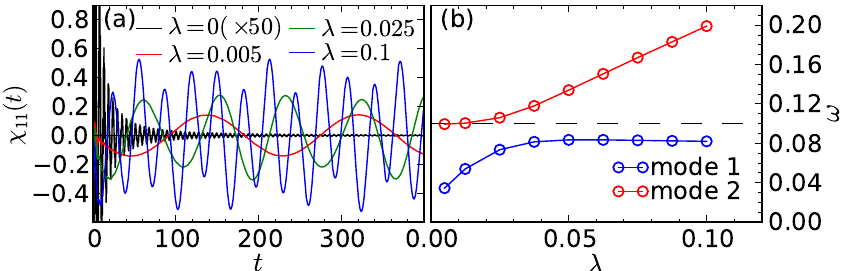} 
  \caption{(a) Susceptibility $\chi^R_{11}(t)$ for different couplings $\lambda$. 
  (b) $\lambda$-dependence of the frequencies of the collective modes at $q=0$ extracted from the peak positions in $-{\rm Im} \chi^R_{11}(\omega)$.
  The dashed black line indicates $\omega_0$.
  }
  \label{fig:fig1}
\end{figure}

In Fig.~\ref{fig:fig1}(a), we show $\chi^R_{11}(t)$ for different phonon couplings. 
Without phonons, $\chi^R_{11}(t)$ oscillates with the same frequency as the minimum gap $\Delta_{\rm EI}=1.15$ and its amplitude decays as $1/t^{1.5}$ (see panel (a)), which is consistent with the previous prediction for strong-coupling SC~\cite{Gurarie2009}. 
(In contrast, the amplitude oscillations in the BCS regime decay as $1/t^{0.5}$~\cite{supplemental}, which is also consistent with the corresponding predictions for SC~\cite{Volkov1974,Yuzbashyan2006a,Barankov2006,Yuzbashyan2006}. Therefore, the existence of prominent amplitude oscllations with frequency of the gap and a decay $\sim 1/t^{0.5}$ may be used to distinguish the BCS from the BEC nature of the system \cite{Lu2017}.)

For $\lambda=0$, the Hamiltonian has a U(1) symmetry and in the EI phase a massless phase mode emerges (the Goldstone mode).
The el-ph coupling breaks the U(1) symmetry and this massless mode becomes massive~\cite{zenker2014}, which leads to additional (undamped) oscillations with two different frequencies in $\chi^R_{11}(t)$, see Fig.~\ref{fig:fig1}(a) and \cite{supplemental}. 
In Fig.~\ref{fig:fig1}(b), we show the dependence of these frequencies on $\lambda$ 
\footnote{The modes have been determined by Fourier transformation of $\chi^R_{11}(t)$ with a damping $\eta=0.006$.}.
The mode whose energy grows from $\omega=0$ corresponds to the massive phase mode. 
(It shows a strong signal in the susceptibility for the phase direction of the order parameter \cite{supplemental}.)
 The mixing between the amplitude and phase oscillations distinguishes EIs from BCS SCs.
The other mode, whose frequency increases from $\omega=\omega_0$ can be regarded as the phonon mode.

Next, we consider the excitation with a laser pulse and discuss how the collective oscillations can be observed and what the conditions for the enhancement of the order are.  We prepare the equilibrium state at $T=0$ at $t=0$ and choose $E(t)=E_0\sin(\Omega t)\exp(-(t-t_{\rm p})^2/(2\sigma_{\rm p}^2))$ with $\Omega=6$, $\sigma_{\rm p}=3$, which corresponds to a $1.56$ eV frequency laser pulse. 
With this choice of parameters, the electrons are directly excited from the lower part of the valence band into the upper part of the conduction band. 
This leads to a substantial Hartree shift (i.e. decrease of $B^z_{k}$) 
due to the change in the band occupations.

 \begin{figure}[t]
  \centering
    \hspace{-0.cm}
    \vspace{0.0cm}
   \includegraphics{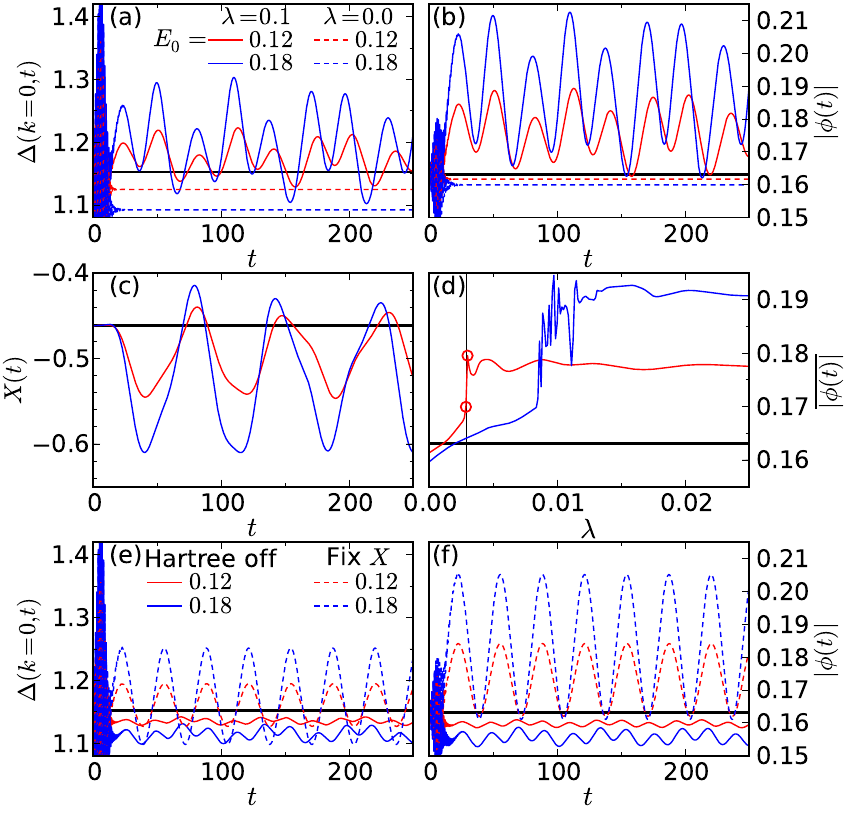} 
  \caption{(a-c) Time evolution of the  gap at $k=0$, $\Delta(k=0,t)$, the excitonic order parameter $|\phi|$, and the phonon displacement  $X$ for various field strengths and el-ph couplings. (d) $\lambda$-dependence of $|\phi|$ averaged over $t\in[0,400]$, for various field strengths. The vertical line indicates the critical value $\lambda_c$ at $E_0=0.12$.
  (e,f) Time evolution of $\Delta(k=0,t)$ and $|\phi|$ for $\lambda=0.1$ evaluated by freezing the Hartree shift (solid lines) and the phonon displacement (dashed lines). We have used $t_p=6.0$. The horizontal black line in each panel shows the corresponding equilibrium values.}
  \label{fig:fig2}
\end{figure}

In Fig.~\ref{fig:fig2}, we show the time evolution of the gap at ${k}=0$ ($\Delta(k=0,t)$), the absolute value of the order parameter ($|\phi|$),  and the phonon displacement ($X$) after the pulse for various conditions \footnote{The gap at each k ($\Delta(k,t)$) is evaluated by diagonalizing the mean-field Hamiltonian at each time, i.e., $\Delta(k,t)=|{\bf B}_k(t)|$. With the present parameters, $\Delta(0,0)=\Delta_{\rm EI}$.}.
When the system is coupled to phonons, we find that $\Delta(k=0,t)$, $|\phi|$ and $|X|$ are enhanced, see Fig.~\ref{fig:fig2}(a-c).  These quantities oscillate with two characteristic frequencies corresponding to the massive phase mode and the phonon mode, as  discussed in the linear response regime. 
With increasing field strength, the enhancement of these quantities becomes more prominent, and the frequencies of the oscillations are slightly increased.
On the other hand, the system without phonons shows a suppression of the gap and order parameter, see Fig.~\ref{fig:fig2}(a-b). 
The dependence of the time average of $|\phi|$ on the el-ph coupling (Fig.~\ref{fig:fig2}(d)) shows that a weak coupling already leads to the enhancement of the order.
At a critical coupling  $\lambda_c$, the system exhibits a dynamical phase transition (discussed below).

 \begin{figure}[t]
  \centering
    \hspace{-0.cm}
    \vspace{0.0cm}
   \includegraphics[width=85mm]{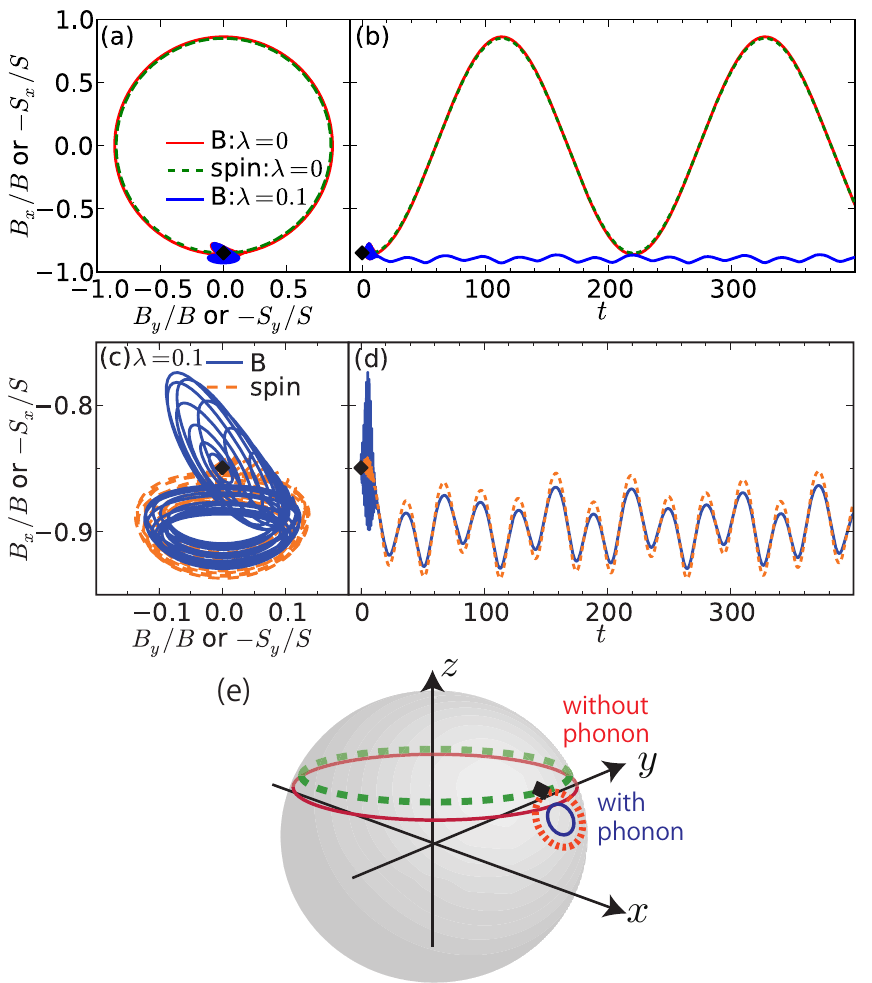} 
  \caption{(a) Trajectory of the normalized pseudo-magnetic field ($B^x/B,B^y/B$) and the normalized pseudo-spin ($-S^x/S,-S^y/S$) at $k=0$  with and without phonons. (b) The time evolution of $B^x/B$ and $-S^x/S$ at $k=0$ with and without phonons. 
  (c) The magnified trajectory of ($B^x/B,B^y/B$) and ($-S^x/S,-S^y/S$) and (d) the time evolution of $B^x/B$ and $-S^x/S$ at $k=0$ for $\lambda=0.1$. The parameters are $t_p=6.0$ and $E_0=0.12$. (e) Schematic picture of the trajectory of ${\bf B}/B$ (solid lines) and $-{\bf S}/S$ (dashed lines) with and without phonons. The equilibrium positions of ${\bf B}/B$ and $-{\bf S}/S$ are indicated by a black diamond in each panel. }
  \label{fig:fig4}
\end{figure}

For the enhancement of the order, the Hartree shift is essential.
If the Hartree shift is artificially suppressed ($B^z$ fixed) in the mean-field dynamics, one finds a
suppression of $\Delta(k=0,t)$ and $|\phi|$,  
see Fig.~\ref{fig:fig2}(e,f). 
After photo-excitation the difference in the occupation, $-\Delta n$, is reduced  
and the bare band gap becomes smaller because of the Hartree shift (a decrease of $B_z$).  In equilibrium, the smaller distance between bands leads to enhanced excitonic condensation.
This argument however assumes thermal distribution functions and cannot be used to explain the transient state. 
The pseudo-spin picture on the other hand suggests an interesting scenario how the change of $B^z$ enhances the order:  
If $B^x$ and $B^y$ would remain static after the sudden decrease of $B^z$ due to the Hartree shift, the pseudo-magnetic field would tilt more to the $x$ direction compared to the equilibrium case. 
This would lead
to a spin precession around the tilted magnetic field, which yields a larger projection on the $x$-$y$ plane, $|S^x_{k}+iS^y_{k}|$, and therefore an enhancement of the order parameter.

In reality, however, the evolution of the pseudo-magnetic field  $B^x$ and $B^y$  and the pseudo-spin must be determined self-consistently. 
In Fig.~\ref{fig:fig4}, we show how they evolve with and without phonons. 
We take ${k}=0$ as a representative
since the region around ${k}=0$ has the largest contribution to the excitonic condensation. For $\lambda=0$, the phase mode is massless, hence after the excitation the magnetic field rotates around the 
$z$-axis 
following the minimum of the free energy. 
The spin follows the magnetic field by keeping the relative angle. 
Hence the spin cannot precess around the magnetic field, and cannot realize the enhancement of $|S^x_{k}+iS^y_{k}|$ from a rotation around a fixed and tilted field mentioned above, see Fig.~\ref{fig:fig4} (a,b,e). 
On the other hand, for $\lambda>0$ the phase mode is massive and therefore the magnetic field is almost confined to the $x$-$z$ plane, see Fig.~\ref{fig:fig4}(a,c).
This means that even in the self-consistent case, {\it when the EI is coupled to phonons}, the result is close to the 
naive static picture  (fixed $B^x$ and $B^y$), 
where the enhancement of the order is explained as a precession around the tilted magnetic field in the $x$ direction. Even though the increase of the projection to the $x$-$y$ plane, $|S^x_{k}+iS^y_{k}|$, after photo-excitation tends to enhance the order, the phases of $S^x_{k}+iS^y_{k}$ for different momenta ${k}$ can lead to destructive interference and a decrease of the order.
However, it turns out that the pseudo-spin at each $k$ roughly follows the magnetic field, see Fig.~\ref{fig:fig4}(d), so that this effect is small. 
We have confirmed that this mechanism, which is based on the massive phase mode, is robust against the frequency of the phonons and the number of phonon branches \cite{supplemental}.

With increasing field strength the oscillations of the magnetic field around the $x$-$z$ plane become large. 
At some critical strength (which depends on $\lambda$), it can overcome the potential barrier of the free energy along the phase direction and starts to rotate around the $z$ axis \cite{supplemental}. 
This can be regarded as a dynamical phase transition (c.f. critical $\lambda_c$ in Fig.~\ref{fig:fig2}(d))\cite{Eckstein2009,Schiro2010,Sciolla2010,Heyl2014,Silva2016}.
Moreover the self-consistent phonon dynamics has an additional positive effect on the enhancement of the order, compare the results with $X$ frozen to the initial value in Fig.~\ref{fig:fig2}(e,f) with Fig.~\ref{fig:fig2}(a,b).
 
 \begin{figure}[t]
  \centering
    \hspace{-0.cm}
    \vspace{0.0cm}
   \includegraphics{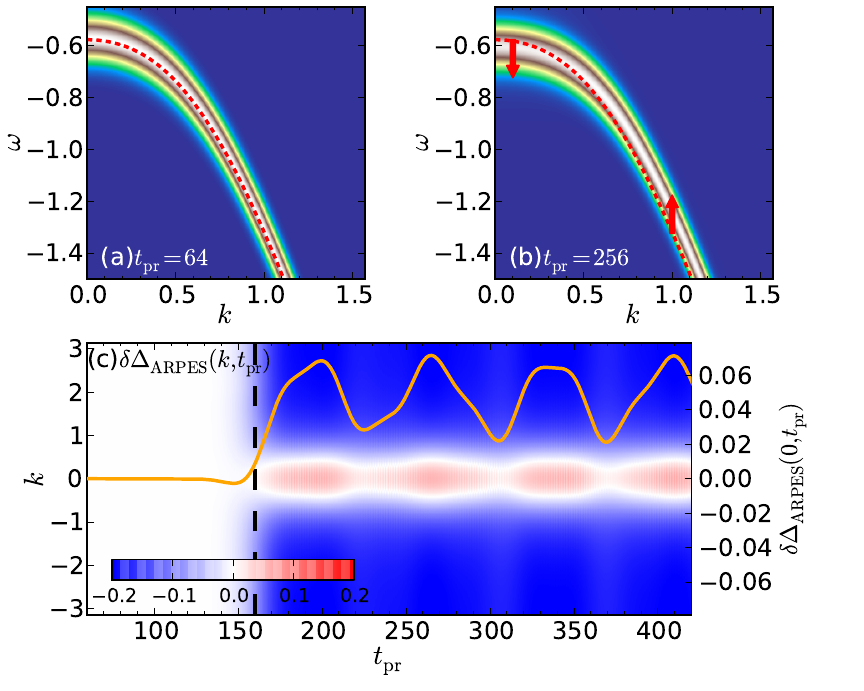} 
  \caption{
  (a)(b) trARPES spectra ($A^R_{k}(\omega;t_{\rm pr})$) before (a) and after (b) the pump. Red dashed lines indicate the equilibrium quasiparticle dispersion from the mean-field theory. (c) Time evolution of the difference between the equilibrium and nonequilibrium gap size at each momentum $k$ ($\delta\Delta_{\rm ARPES}(k,t_{\rm pr})$).  The orange solid line is $\delta\Delta_{\rm ARPES}(0,t_{\rm pr})$. 
  Parameters are $\lambda=0.1$, $E_0=0.18$, $t_{\rm p}=160.0$, and $\sigma_{\rm pr}=12.0$. 
  The black dashed line indicates the center of the pump pulse.
 }
  \label{fig:fig3}
\end{figure}
Finally, we discuss how the enhancement of the excitonic order and gap can be observed in experiments. In Fig.~\ref{fig:fig3}(a,b), we show the trARPES spectrum evaluated as \cite{Freericks2009,eckstein2008} $A^R_{k}(\omega;t_{\rm pr})=-\frac{1}{\pi} {\rm Im} \int\int dt dt' s(t-t_{\rm pr})s(t'-t_{\rm pr})e^{i\omega(t-t')}[G_{{k},00}^R(t,t')+G_{{k},11}^R(t,t')]/2$
before and after the laser pump. 
Here $G_{{k},\alpha\beta}^R(t,t')=-i\theta(t-t')\langle[\hat{c}_{k,\alpha}(t),\hat{c}^\dagger_{k,\beta}(t')]_{+}\rangle$ is the retarded part of the Green's function, $s(t)=\frac{1}{\sqrt{2\pi}\sigma_{\rm pr}}\exp(-t^2/(2\sigma^2_{\rm pr}))$, $t_{\rm pr}$ is the probe time, and $\sigma_{\rm pr}$ is the width of the probe pulse.
Before the pump, the peak in the spectrum follows the quasi-particle dispersion from the mean-field theory in equilibrium.
After the pump, the weight of the spectrum shifts away from the Fermi level around ${k}={0}$ (the $\Gamma$ point), while 
away from ${k}={0}$, it shifts closer to the Fermi level.
In Fig.~\ref{fig:fig3}(c), we show the time evolution of the difference between the equilibrium and nonequilibrium band distance at each ${k}$ ($\delta\Delta_{\rm ARPES}(k,t_{\rm pr})$), which is determined from the difference between the peaks in $A^R_{k}(\omega;t_{\rm pr})$. 
The shift in the band distance oscillates with the frequencies observed in the analysis of $\chi_{11}$ (the orange line in Fig.~\ref{fig:fig3}(c)), which demonstrates that the collective modes can be observed with trARPES.

{\it Conclusions--}
We have revealed that the el-ph coupling, which is associated with the structural transition in ${\rm  Ta_2 Ni Se_5}$, has a large and qualitative effect  
on the dynamics of the excitonic insulator.
In particular, we demonstrated a novel mechanism for photo-enhanced excitonic order based on a cooperative effect between the el-ph coupling, the massive phase mode and the Hartree shift.
Combining photo excitation and a reduction of the symmetry of the Hamiltonian may provide a new strategy to enhance analogous orders such as SC\cite{Hoshino2015}.
Although the mean-field dynamics ignores scattering processes and therefore the long time dynamics is not reliable, it can capture the short time dynamics~\cite{Tsuji2013}. 
Since the enhancement observed here occurs quickly after the pump, we expect it to be a genuine effect.

The proposed mechanism for photo-enhanced excitonic order is an alternative to the one discussed in Ref.~\onlinecite{Mor2016}, where the enhancement was attributed to nonthermal distribution functions with additional Hartree shifts from higher conduction bands. 
Both mechanisms are consistent with the experimental results reported in Ref.~\onlinecite{Mor2016}, which show a gap enhancement around the $\Gamma$ point and a reduction away from it.
We believe that they could be combined; e.g., at short times the drastic effects of the el-ph coupling and the additional Hartree shifts
cooperate, while the situation considered in Ref.~\onlinecite{Mor2016} is relevant at later times. 
The combined study of the evolution of the phonons, the order parameter, and the nonthermal distribution functions (including incoherent collision processes) is an 
interesting topic for future work.

\acknowledgments
 The authors wish to thank C.~Monney, S.~Mor, M.~Sch\"{u}ler and J.~Kune\v s for fruitful discussions.
YM and DG are supported by the Swiss National Science Foundation through NCCR Marvel and Grant No. 200021-165539. 
ME acknowledges support by the Deutsche Forschungsgemeinschaft within the Sonderforschungsbereich 925 (projects B4).
PW acknowledges support from ERC Consolidator Grant No.~724103.

\bibliographystyle{prsty}
\bibliography{Ref}

\clearpage
\appendix

\subsection{Effects of other phonon modes}
In the main text, we have considered a single phonon branch which is coupled to the transfer integral between the electron bands.
In a more realistic scenario, the system is coupled to multiple phonon branches, and the question we want to address here is whether or not this has a qualitative effect on the mechanism discussed in the main text.
In the following, we first introduce another type of electron-phonon (el-ph) coupling, which has been pointed out in the recent LDA calculation of  Ref.~\cite{Kaiser2016}.
Then we demonstrate that, in order to see the enhancement of the EI order and the gap, one needs at least one phonon mode that 
is coupled to the transfer integral between the electron bands. Hence, the number of phonon modes does not influence the physics we discussed in the main text.

Based on LDA calculations in Ref.~\cite{Kaiser2016}, it was pointed out that a phonon mode at 1THz modifies the hybridization between bands as well as the on-site energy (crystal field splitting).
If a phonon mode only has the latter effect,
its Hamiltonian can be described by 
\begin{align}
&H_{\rm el-ph,2}+\hat{H}_{\text{ph},2}\nonumber\\
&=g_2\sum_i (\hat{b}_{i,2}^\dagger+\hat{b}_{i,2})(\hat{n}_{i,0}-\hat{n}_{i,1})+\omega_2\sum_i\hat{b}_{i,2}^\dagger \hat{b}_{i,2}.\label{eq:type2}
\end{align}
We denote the corresponding creation operator by $\hat{b}_{i,2}^\dagger$, the coupling constant by $g_2$, the phonon frequency by $\omega_2$, 
and we introduce $\lambda_2\equiv2g_2^2/\omega_2$. 
This Hamiltonian represents optical phonons coupled to the electrons through 
the difference in occupancy between the valence and conduction bands, which is a different type of el-ph coupling than the one considered in the main part.
In the following, we denote the el-ph coupling in the main text as ``type 1" and the one introduced above as ``type 2".
We note that in general a given phonon mode can exhibit these two types of couplings at the same time, 
but for simplicity in this study we assume that each phonon mode possesses only one of them. 

The dynamics of the type 2 phonons can also be treated within the mean-field theory by introducing the mean phonon displacement 
$X_2(t)\equiv \langle \hat{b}_{i,2}^\dagger(t)+\hat{b}_{i,2}(t)\rangle$.
The decoupling of the interaction term is discussed in the next section. 
The dynamics of the pseudo-spins (electrons) is described 
by Eq.~(2) 
of the main text, where $B_{k}^x$ and $B_{k}^y$ are given by Eq.~(3a) and Eq.~(3b), respectively, and 
\begin{align}
B_{k}^z(t)&=2\epsilon_{k}^z-U\Delta n(t)+2g_2X_2(t).
\end{align}
The equation of motion for the phonons is $\partial_t P_2(t)=-\omega_2X_2(t)-2g_2\Delta n(t)$ and  $\partial_t X_2(t)=\omega_2 P_2(t)$.

Let us first note that we cannot expect the enhancement of the EI order with the second type of el-ph coupling only.
In the mechanism explained in the main part, a cooperative effect between the massive phase mode and the Hartree shift 
was essential for the enhancement.
However, the second type of el-ph coupling does not break the U(1) symmetry, thus the EI breaks the continuous symmetry and the phase mode remains massless.
Therefore, the mechanism discussed in the main text does not work, which is numerically shown below.

In Fig.~\ref{fig:supp_fig1}, we show the properties of collective amplitude oscillations and the photo-induced dynamics for the case with only the second type of phonons.
Here we use as a reference the same parameters as in the main text, i.e. $\Delta_0=-0.55$, $\Delta_1=-2.45$, $\lambda_2=0$ and $U=3$.
We take $\omega_2=0.1$ and, for the pump pulse, $E(t)=E_0\sin(\Omega t)\exp(-(t-t_{\rm p})^2/(2\sigma_{\rm p}^2))$ with $\Omega=6$, $\sigma_{\rm p}=3$ and $t_{\rm p}=6$.
In Fig.~\ref{fig:supp_fig1}(a), we show $\chi_{11}^R(t)$. 
As in the case without phonons, this quantity oscillates with $\Delta_{\rm EI}$ and
its amplitude decays $\sim 1/t^{1.5}$.
In Figs.~\ref{fig:supp_fig1}(b-d), we show the time evolution of 
several observables after the pump pulse.
The gap at $k=0$ ($\Delta(k=0,t)$), 
the excitonic order parameter $(|\phi|$), and the phonon displacement ($X_2$) 
are suppressed after the pulse.
As we increase the pulse 
amplitude 
the suppression becomes larger.
The gap shows oscillations with the frequency $\omega_2$, which 
originate 
from the phonon dynamics,
while the value of the order parameter is almost constant.

 \begin{figure}[t]
  \centering
    \hspace{-0.6cm}
    \vspace{0.0cm}
       \includegraphics{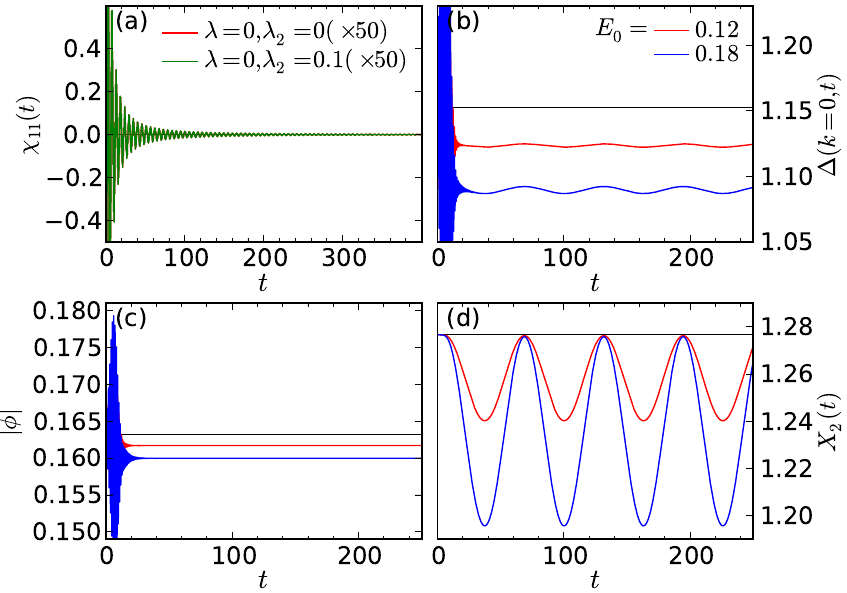} 
  \caption{(a) Susceptibility  ($\chi^R_{11}(t)$) without phonons and with the type 2 phonons.
  The two curves are on top of each other. (b-d) 
  Time evolution of the band gap at $k=0$ 
 ($\Delta(k=0,t)$),
  the excitonic order parameter ($|\phi|$), and the phonon displacement ($X_2$) for various field strengths for  $\lambda=0,\lambda_2=0.1$.}
  \label{fig:supp_fig1}
\end{figure}

Now we discuss the effects of multiple phonon branches.
First we note that in reality we do not need to consider many phonon branches,
since in the experiment of Ref.~\cite{Kaiser2016}, only three prominent modes at 1, 2, and 3 THz were observed, with a strong signal from the 1THz and 3THz modes and a weaker signal from the 2THz mode.
Motivated by this, we only consider two phonon branches and set the phonon frequencies to mimic the 1 THz and 3 THz modes,
which corresponds to $\omega_{0,a}=0.015$ and $\omega_{0,b}=0.045$, respectively.
Unfortunately, the detailed properties of the el-ph couplings are not available in the literature.
Therefore, we assumed that they are either of type 1 or type 2.  
We thus checked three cases, i.e. (type 1, type 1), (type 1, type 2) and (type 2, type 1).
In all cases, we have confirmed that the enhancement of the EI order and the gap can be observed in a manner analogous to the single phonon branch set-up studied in the main text.

Here as a representative of these three cases we show the results of the case where the $\omega_{0,a}$ phonon is of type 1, while the $\omega_{0,b}$ phonon is of type 2.
The Hamiltonian that includes these phonons explicitly reads 
\begin{align}
&H_{\rm el-ph}+H_{\rm ph}\nonumber\\
&=g_a\sum_i (\hat{b}_{i,a}^\dagger+\hat{b}_{i,a})(\hat{c}^\dagger_{i,1}\hat{c}_{i,0}+\hat{c}^\dagger_{i,0}\hat{c}_{i,1})+\omega_{0,a}\sum_i \hat{b}_{i,a}^\dagger \hat{b}_{i,a}\nonumber\\
&+g_b\sum_i (\hat{b}_{i,b}^\dagger+\hat{b}_{i,b})(\hat{n}_{i,0}-\hat{n}_{i,1})+\omega_{0,b}\sum_i \hat{b}_{i,b}^\dagger \hat{b}_{i,b},\label{eq:2phonons}
\end{align}
where $\hat{b}_{i,\gamma}$ is the annihilation operator for the phonon branch $\gamma$.
We also introduce $\lambda_\gamma=2g_\gamma^2/\omega_{0,\gamma}$.
The electron part of the Hamiltonian is the same as in Eq.~(1) of the main text.

\begin{figure}[t]
  \centering
    \hspace{.0cm}
    \vspace{0.0cm}
   \includegraphics{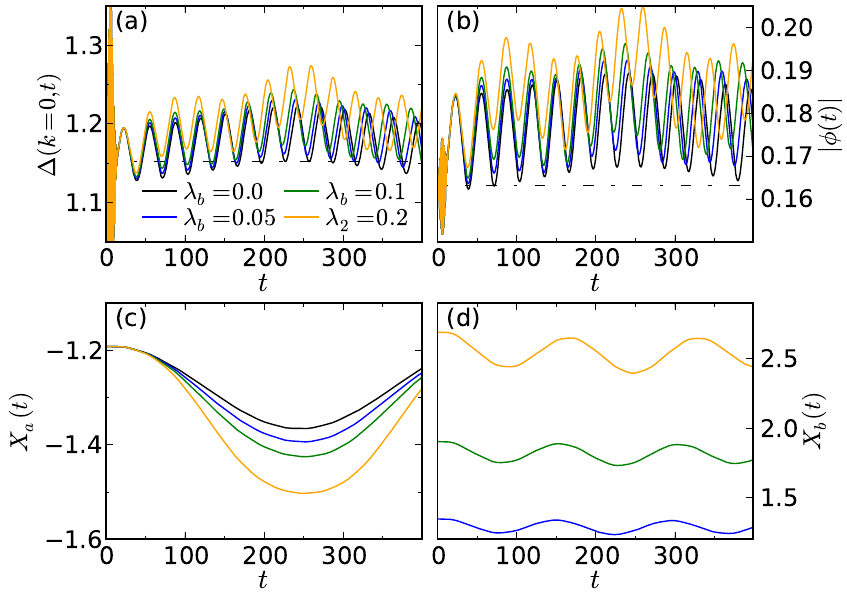} 
  \caption{ Time evolution of the excitonic insulator with two phonon branches, Eq.~(\ref{eq:2phonons}).
  (a-d) Time evolution of the gap at $k=0$, $\Delta(k=0,t)$, the excitonic order parameter $|\phi|$, and the phonon displacement $X_a$ and $X_b$ for $\omega_{0,a}=0.015,\;\omega_{0,b}=0.045,\;\lambda_a=0.1$, $E_0=0.12$, $T=0$ and various $\lambda_b$ (coupling strength of the type 2 phonon).  We use $\Delta_0=-0.55$, $\Delta_1=-2.45$, $\lambda=0$ and $U=3$ as a reference as is discussed in the main text. The pump condition is the same as in the main text, $\Omega=6$, $\sigma_{\rm p}=3$.}
  \label{fig:resp_fig2_dum}
\end{figure}

In Fig.~\ref{fig:resp_fig2_dum}, we show the results for various sets of el-ph couplings.
In all cases, we can see an enhancement of the EI gap and order parameter and the dynamics of $X_a(t)$ and $X_b(t)$ show that the 
main oscillation component is coming from $\omega_{0,a}$ and $\omega_{0,b}$, respectively.
These results indicate that the frequency of the type 1 phonon does not affect the enhancement of the order
(note that the frequency of the type 1 phonon used here is much smaller than the one used in the main text)
and that the number of phonon branches is also irrelevant as far as at least one phonon is of type 1.
One interesting observation here is that as we increase the coupling of the type 2 phonon, the enhancement of the EI gap and the EI order 
becomes larger. 
This can be explained as a positive feedback effect from the type 2 phonons, similar to what we observed for the type 1 phonon in the main text.
Since the phonon can move it can adjust its position to a more preferable point.  
In this case, because of the photo-doping, $|\Delta n(t)|$  first decreases and then the size of phonon displacement ($|X|$) becomes smaller.
Because this phonon mode couples to the potential on each site, this yields a further reduction of the $z$ component of the pseudo-magnetic field (${\rm B}$ is more tilted towards the x axis.).
This can further enhance the EI order.

Before we finish this section, we add some comments on 1) the effects of a phonon damping term and  2) the temperature dependence of the dynamics.
To investigate point 1), we have introduced a damping term in the mean-field phonon dynamics, i.e. $\partial_t P(t) = -2\xi \omega_0 P(t)-\omega_0 X(t)-4g{\rm Re} \phi(t)$ with $\xi$ the damping factor.
These calculations confirmed that even with a strong damping the enhancement of the order is realized after the pump. 
(It is important to note that the experiment shows that the oscillations after the pump are long-lived, hence in practice we do not need to worry about the phonon damping effects on the physics discussed in this paper.)
As for 2), we have repeated the same calculations as in the main text for different initial temperatures. 
It turns out that, without the el-ph coupling, the order parameter and the gap decrease after the pulse at all temperatures,
while with the el-ph coupling one can observe an enhancement up to temperatures very close to the thermal transition.
Above the transition temperature,  both cases show no enhancement of  the order parameter and a suppression of the gap.

\subsection{Mean-field theory}
In the mean-field theory, at each time $t$, we decouple the interaction term as 
\begin{align}
&\hat{n}_{i,0}\hat{n}_{i,1}\longrightarrow  \\
&\langle n_{i,0}(t)\rangle \hat{n}_{i,1}+\langle n_{i,1}(t)\rangle \hat{n}_{i,0}-\phi(t) \hat{c}_{i,1}^\dagger \hat{c}_{i,0}-\phi(t)^*\hat{c}_{i,0}^\dagger \hat{c}_{i1}\nonumber,
\end{align}
 where the first two terms correspond to the Hartree terms and the latter two to the Fock terms. We also decouple the el-ph coupling terms as 
 \begin{subequations}
 \begin{align}
 &(\hat{b}_i^\dagger+\hat{b}_i)(\hat{c}^\dagger_{i,1}c_{i,0}+\hat{c}^\dagger_{i,0}c_{i,1})\nonumber\\
 &\rightarrow X(t)(\hat{c}^\dagger_{i,1}\hat{c}_{i,0}+\hat{c}^\dagger_{i,0}\hat{c}_{i,1}) +2{\rm Re}\phi_0(t)(\hat{b}_i^\dagger+\hat{b}_i),\\
 & (\hat{b}_{i,2}^\dagger+\hat{b}_{i,2})(\hat{n}_{i,0}-\hat{n}_{i,1})\nonumber\\
  &\rightarrow X_2(t) (\hat{n}_{i,0}-\hat{n}_{i,1})+(\hat{b}_{i,2}^\dagger+\hat{b}_{i,2}) \Delta n(t).
\end{align}
\end{subequations}
This decoupling is applicable for any systems with phonon branches of the type 1 or the type 2.
In the following, we focus on the case where the electron part is coupled to one type 1 phonon branch and one type 2 phonon branch.
Namely, the total Hamiltonian of the system is the sum of Eq.~(1) in the main text and the above Eq.~(\ref{eq:type2}).

This leads to the mean-field Hamiltonians
\begin{subequations}
\begin{align}
H_{\rm el}^{MF}(t)&=\frac{1}{2}\sum_k \hat{\Psi}_k^\dagger
\begin{bmatrix}
C^0_k+B^z_k(t) & B^x_k(t)-iB^y_k(t)\\
 B^x_k(t)+iB^y_k(t) & C^0_k-B^z_k(t)
\end{bmatrix}
 \hat{\Psi}_k\nonumber\\
&=\sum_{\rm k} {\bf B}_{k}(t)\cdot {\hat{\bf S}}_{k}+C^0_{k}\cdot \hat{I}_{k},\label{eq:MF_ele}\\
H^{\rm MF}_{\rm ph0}(t)&=\omega_0\sum_i\hat{b}^\dagger_i\hat{b}_i+g(\phi(t)+\phi(t)^*)\sum_i  \hat{X}_i,\\
H^{\rm MF}_{\rm ph2}(t)&=\omega_2\sum_i \hat{b}^\dagger_{i,2} \hat{b}_{i,2}+g_2\Delta n(t)\sum_i  \hat{X}_{i,2},
\end{align}
\end{subequations}
where $\hat{X}_i=\hat{b}_i+\hat{b}^\dagger_i$, $\hat{X}_{i,2}=\hat{b}_{i,2}+\hat{b}^\dagger_{i,2}$ and
\begin{align}
C_{\rm k}^0&=(\epsilon_{{k},0}+\epsilon_{{k},1})+(\Delta_0+\Delta_1)+U (n_0+ n_1).
\end{align}
Here $n_\alpha$ indicates the number of electrons per site  in the $\alpha$ band and $n_0+ n_1$ is constant.
The equations of motion for observables shown in the main text are obtained from these mean-field Hamiltonians.
By diagonalizing Eq.~(\ref{eq:MF_ele}) at each time, one can obtain the time-dependent (instantaneous) dispersion of the electrons, $E_{\pm}(k,t)\equiv(\pm |{\rm B}_k(t)|+C^0_k)/2$. Then the gap at each $k$ ($\Delta (k,t)$) becomes $|{\rm B}_k(t)|$.

In equilibrium the mean-field theory yields the following conditions. For the type 1 phonons,
\begin{align}
P&=0,\,\,\,\,\,X=-\frac{4g}{\omega_0}{\rm Re}\phi,
\end{align}
 and,
 for the type 2 phonons, 
\begin{align}
P_2&=0,\,\,\,\,\,X_2=-\frac{2g_2}{\omega_2}\Delta n.
\end{align}
Here we note that since we are mainly interested in the electron dynamics, we only need the information on the average phonon displacement in the mean-field description. General quantities such as phonon occupations still depend on temperature, but here we do not need this information.

As for the electrons, by diagonalizing the mean-field Hamiltonian, Eq.~(\ref{eq:MF_ele}), and evaluating the expectation values of physical quantities from the thermally occupied eigenstates,
we obtain the self-consistency relation 
\begin{subequations}
\begin{align}
&\phi=\frac{1}{N}\sum_{k}\frac{B_{k}^x+i B_{k}^y}{2 B_{k}}
[f(E_{+}(k),T)-f(E_{-}(k),T)],\\
&\Delta n=\frac{1}{N}\sum_{k} \frac{B_{k}^z}{B_{k}}
[f(E_{+}(k),T)-f(E_{-}(k),T)],\\
&n_0+ n_1=\frac{1}{N}\sum_{k} [f(E_{+}(k),T)+f(E_{-}(k),T)].
\end{align}
\end{subequations}
Here the components of the pseudo-magnetic field are
\begin{subequations}
\begin{align}
B_{k}^x&=-2(U+2\lambda){\rm Re} \phi,\\
B_{k}^y&=-2U{\rm Im} \phi,\\
B_{k}^z&=2\epsilon_{{k},z}-\left(U+2\lambda_2\right) \Delta n,\\
\label{dwbicbds}
B_{k}&=\sqrt{( B_{\rm k}^{x})^2+( B_{\rm k}^{y})^2+( B_{\rm k}^{z})^2},
\end{align}
\end{subequations}
and $E_{\pm}(k)\equiv(\pm B_k+C^0_k)/2$.
The Fermi distribution function at the temperature $T$ is $f(\epsilon,T)=1/(e^{\epsilon/T}+1)$. 
Without the type 1 el-ph coupling ($g=0$), we can choose an arbitrary phase of the order parameter $\phi$, while for $g\neq0$, the order parameter $\phi$ becomes real and the remaining degree of freedom is its sign. 

We also show the expression for the Green's functions since we use in the next section.
The lesser and greater parts of the Green's functions are defined as $G^<_{k,\alpha,\beta}(t,t')\equiv i\langle \hat{c}^\dagger_{k,\beta}(t') \hat{c}_{k,\alpha}(t)\rangle$ and $G^>_{k,\alpha,\beta}(t,t')\equiv -i\langle \hat{c}_{k,\alpha}(t) \hat{c}^\dagger_{k,\beta}(t')\rangle$ and we can regard them as $2\times2$ matrices in terms of the band index.
In equilibrium within the mean-field theory with real order parameter, they are expressed as 
\begin{subequations}\label{eq:G_eq}
\begin{align}
\hat{G}^{<}_k(t) &=\frac{i}{2} \sum_{\alpha=\pm}  f(E_{\alpha}(k),T)e^{-iE_{\alpha}(k)t}\hat{M}_{\alpha}(k),\\
\hat{G}^{>}_k(t)&= -\frac{i}{2} \sum_{\alpha=\pm} (1-  f(E_{\alpha}(k),T))e^{-iE_{\alpha}(k)t}\hat{M}_{\alpha}(k),
\end{align}
\end{subequations}
where 
\begin{align}
\hat{M}_{\pm}(k)&=\pm\frac{B^x_k}{B_k}\sigx\pm\frac{B^z_k}{B_k}\sigz+\hat{\sigma}_0. 
\end{align}

\subsection{Additional study of susceptibilities}
The dynamical susceptibilities evaluated from the mean-field dynamics correspond to those evaluated within the random phase approximation (RPA).
In this section, we discuss this point in detail and show additional results for the susceptibility for the phase direction of the excitonic order parameter.

First, we consider the following four types of external homogeneous perturbations
\begin{align}\label{eq:small_field}
\hat{H}_{{\rm ex},\nu} (t)=\delta F_{{\rm ex},\nu} (t) \sum_{k} \hat{\Psi}_k^\dagger \hat{\sigma}_\nu \hat{\Psi}_k,
\end{align}
where $\nu=0,1,2,3$ and $\hat{\sigma}_\nu$ denotes the Pauli matrix.
We also introduce $\hat{\rho}_\mu\equiv\frac{1}{N}\sum_{k} \hat{\Psi}_k^\dagger \hat{\sigma}_\mu \hat{\Psi}_k$.
In the linear response regime, we can define the (full) susceptibility, $\chi^R_{\mu\nu} (t,\bar{t})$, as 
\begin{align}\label{eq:chi}
\delta\langle \hat{\rho}_{\mu}(t)\rangle&=\sum_\nu \int d\bar{t} \chi^R_{\mu\nu} (t,\bar{t}) \delta F_{{\rm ex},\nu} (\bar{t}).
\end{align}
$\chi^R_{\mu\nu}$ detects collective modes with zero momentum.
When the order parameter $\phi$ is taken real, $\langle \hat{\rho}_{1}(t)\rangle$ denotes the dynamics along the amplitude direction, while 
$\langle \hat{\rho}_{2}(t)\rangle$ corresponds to that along the phase direction of the order parameter.
Therefore, if there is no mixing between the amplitude and the phase, $\chi^R_{11}$ and $\chi^R_{22}$ can be used to detect the amplitude mode and the phase mode, respectively \cite{Kulik1981}.

Now we consider the expression of $\chi^R_{\mu\nu}$, which corresponds to the mean-field dynamics.
We can rewrite the mean-field Hamiltonians as 
\begin{subequations}\label{eq:H_MF2}
\begin{align}
H^{\rm MF}_{\rm el}(t)&=H^{\rm MF}_{\rm el,eq}+\sum_{i,\nu} (\delta F_{{\rm ex},\nu} (t)+\delta F_{\nu} (t)) \hat{\rho}_\nu,\\
H^{\rm MF}_{\rm ph0}(t)&=H^{\rm MF}_{\rm ph0,eq}+\delta H_{0} (t) \sum_{i} \hat{X}_i,\\
H^{\rm MF}_{\rm ph2}(t)&=H^{\rm MF}_{\rm ph2,eq}+\delta H_{2} (t) \sum_{i} \hat{X}_{i,2}.
\end{align}
\end{subequations}
Here $H^{\rm MF}_{\rm eq}$ represents the mean-field Hamiltonians in equilibrium.
$\delta F_{\nu} (t),\delta H_{0} (t)$ and $\delta H_{2} (t)$ are the changes in the mean-fields relative to the equilibrium values,
\begin{subequations}\label{eq:dMF}
\begin{align}
\delta F_{\nu} (t)&=U_{\nu} \delta \langle \hat{\rho}_{\nu}(t)\rangle+\delta_{\nu,1}g  \delta X(t) +\delta_{\nu,3} g_2 \delta X_2(t),\\
\delta H_{0} (t)&=g\delta \langle \hat{\rho}_{1}(t)\rangle,\\
\delta H_{2} (t)&=g_2\delta \langle \hat{\rho}_{3}(t)\rangle,
\end{align}
\end{subequations}
with $[U_0,U_1,U_2,U_3]=[-U/2,U/2,U/2,U/2]$.
Here $\delta \langle \hat{\rho}_{\nu}(t)\rangle$, $\delta X(t)$ and $\delta X_2(t)$ denote the difference from the equilibrium values.

In the linear response regime, from the standard Kubo formula and Eq.~(\ref{eq:H_MF2}),
\begin{subequations}\label{eq:MF_Kubo1}
\begin{align}
\delta\langle \hat{\rho}_{\mu}(t)\rangle&=\sum_\nu \int d\bar{t} \chi^R_{0,\mu\nu} (t,\bar{t}) [ \delta F_{{\rm ex},\nu} (\bar{t})+\delta F_{\nu} (\bar{t})],\\
\delta X_{0}(t)&=\int d\bar{t} D^R_{0} (t,\bar{t}) \delta H_{0} (\bar{t}),\\
\delta X_{2}(t)&=\int d\bar{t} D^R_{2} (t,\bar{t}) \delta H_{2} (\bar{t}).
\end{align}
\end{subequations}
Here, $\chi^R_{0}$ is the susceptibility computed with the mean-field Hamiltonian with the mean-field fixed to the equilibrium value.
It corresponds to bubble diagrams in the language of Feynman diagrams,
\begin{align}
\chi_{0,\mu\nu}^R(t)&=-i\theta(t)\frac{1}{N}\sum_k  \Bigl\{{\rm tr} [\hat{\sigma}_\mu \hat{G}^>_{k}(t) \hat{\sigma}_{\nu} \hat{G}^<_k(-t)\nonumber\\
&-{\rm tr} [\hat{\sigma}_\mu \hat{G}^<_{k}(t) \hat{\sigma}_{\nu} \hat{G}^>_k(-t) ]\Bigl\}.
\end{align}
 $D^R_0(t,t')\equiv-i\theta(t-t')\langle [\hat{X}_0(t),\hat{X}_0(t')]\rangle_0$ and  $D_2(t,t')\equiv-i \theta(t-t')\langle [\hat{X}_2(t),\hat{X}_2(t')]\rangle_0$ are the retarded parts of the free phonon Green's functions for the type 1 and type 2 phonons, respectively. 
 $\theta(t-t')$ is the Heaviside step function.
By substituting Eq.~(\ref{eq:dMF}) into Eq.~(\ref{eq:MF_Kubo1}), we obtain the self-consistent equation for $\delta\langle \hat{\rho}_{\nu}(t)\rangle$ expressed with $\chi^R_{0,\mu\nu}, D^R_0$ and $D^R_2$. We compare it with Eq.~(\ref{eq:chi}) and apply the Fourier transformation.
This leads to a $4\times4$ system of equations for the susceptibility $\chi_{\mu\nu}$ in the following form
\begin{align}
\hat{\chi}^R(\omega)=\hat{\chi}^R_0(\omega)+\hat{\chi}_0^R(\omega)\hat{\Theta}(\omega)\hat{\chi}^R(\omega),\label{eq:chi_dyson}
\end{align}
where we have identified the irreducible vertex part
\begin{align}
\hat{\Theta}(\omega)=
\begin{bmatrix}
\frac{U}{2} & 0 & 0 & 0\\
0 & -\frac{U}{2}+g^2 D_0^R(\omega) & 0 & 0\\
0 & 0 & -\frac{U}{2} & 0\\
0 & 0& 0 & -\frac{U}{2}+g^2_2 D_2^R(\omega)
\end{bmatrix}.\label{eq:chi_dyson2}
\end{align}
Here we note that the components $\chi^R_{0\nu}$ are  zero since no perturbation considered here 
changes
the total number of electrons. The external field proportional to the total number of electrons does not alter the dynamics, since it commutes with the Hamiltonian, and therefore $\chi^R_{\mu 0}$ are zero. 
From this consideration one can also see that the bare susceptibilities $\chi^R_{0, 0\nu}$ and $\chi^R_{0,\mu 0}$ are zero. 
Therefore, in practice, we only need to focus on $\mu,\nu=1,2,3$ in Eq.~(\ref{eq:chi_dyson}) and Eq.~(\ref{eq:chi_dyson2}).

From Eq.~(\ref{eq:G_eq}), the bare susceptibility is given by
\begin{align}
\chi_{0,\mu\nu}^R(t)\nonumber=&\theta(t)(-i\frac{1}{N})\sum_{k}\sum_{a=\pm} \frac{f(E_{\bar{a}}(k))-f(E_{a}(k))}{4}\\
&\times e^{-i(E_{a}(k)-E_{\bar{a}}(k))t} 
{\rm tr}[\hat{\sigma}_\mu \hat{M}_a \hat{\sigma}_\nu\hat{M}_{\bar{a}}] 
\end{align}
and its Fourier transform yields a generalization of the Linhard formula
\begin{align*}
  &\hat{\chi}_{0}^R(\omega)=\frac{1}{N}\sum_k\frac{f(E_-(k),T)-f(E_+(k),T)}{4}\\
  &\times \left(\frac{\hat{A}_k}{\omega^+-(E_+(k)-E_-(k))}-\frac{\hat{A}_k^{T}}{\omega^++(E_+(k)-E_-(k))}\right),
\end{align*}
where 
$\omega^+=\omega+i\eta$, and the matrices $\hat{A}_k$ are obtained by the evaluation of ${\rm tr}[\hat{\sigma}_\mu \hat{M}_a \hat{\sigma}_\nu\hat{M}_{\bar{a}}]$:
\begin{align}
  \hat{A}_{k}=4
  \begin{bmatrix}
  (\frac{B_k^z}{B_k})^2 & -i\frac{B_k^z}{B_k} & \frac{B_k^x B_k^z}{B_k^2}\\
  i\frac{B_k^z}{B_k} &1 &i \frac{B_k^x}{B_k}\\
  \frac{B_k^x B_k^z}{B_k^2} & -i \frac{B_k^x}{B_k} & (\frac{B_k^x}{B_k})^2
  \end{bmatrix}.
\end{align}
From these expressions, one can see that the amplitude oscillations (represented by the 11 component) are coupled to the phase oscillations. 
For example, at $T=0$ and if $E_+(k)>0$ and $E_-(k)<0$ for all $k$, which is the case for the parameters used in the main text,
\begin{subequations}
\begin{align}
\chi_{0,12}^R(t)&=\theta(t)\frac{-2}{N}\sum_k \frac{B_k^z}{B_k}\cos(B_kt),\\
\chi_{0,12}^R(\omega)&=\frac{-2i}{N}\sum_k \frac{B_k^z}{B_k}\frac{\omega^+}{(\omega^+)^2-(E_+(k)-E_-(k))^2}.
\end{align}
\end{subequations}
Since $B_k^z/B_k$ is always positive, this term does not vanish, which leads to the mixing between amplitude and phase oscillations.
This is in sharp contrast to the case of BCS superconductors, where the amplitude oscillations are decoupled from other components \cite{Kulik1981}.

\begin{figure}[t]
  \centering
    \hspace{-0.cm}
    \vspace{0.0cm}
   \includegraphics{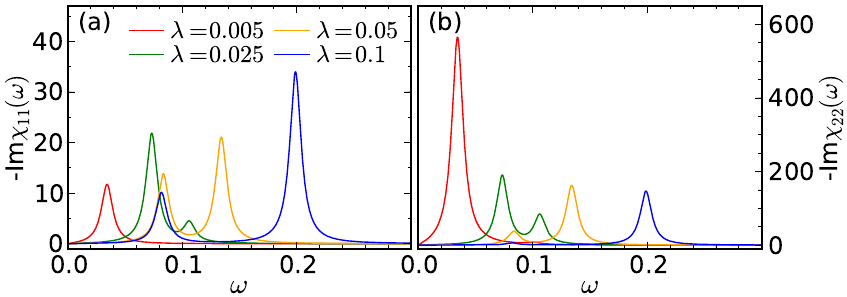} 
  \caption{(a)  Imaginary part of the susceptibility for the amplitude direction of the excitonic order parameter ($-{\rm Im} \chi^R_{11}(\omega)$).
  (b) Imaginary part of the susceptibility for the phase direction of the excitonic order parameter ($-\text{Im}\chi^R_{22}(\omega)$). 
   We use the same parameters as in the main text. 
   The frequency dependent quantities $-\text{Im}\chi^R(\omega)$ are numerically obtained from a Fourier transformation of $\chi^R_{\mu\mu}(t)$ with a damping factor $e^{-\eta t}$ and $\eta=0.006$.
   }
  \label{fig:supp_fig2}
\end{figure}

In Fig.~\ref{fig:supp_fig2}, we compare  the imaginary parts of the susceptibilities for the amplitude direction and the phase direction of the excitonic order parameter ($-\text{Im}\chi^R_{11}(\omega)$ and $-\text{Im}\chi^R_{22}(\omega)$). Here $\chi^R_{\mu\mu}(\omega)$ is obtained by the Fourier transformation of $\chi^R_{\mu\mu}(t)$, which we directly measure by putting a field as defined in Eq.~(\ref{eq:small_field}).
We can see that the peaks in $-\text{Im}\chi^R_{11}(\omega)$ and $-\text{Im}\chi^R_{22}(\omega)$ emerge at the same positions.
This means that the amplitude and phase oscillations are coupled.
However, the mode which emerges from zero as we increase $\lambda$ has larger intensity in $-\text{Im}\chi^R_{22}(\omega)$ than in $-\text{Im}\chi^R_{11}(\omega)$.
This indicates that this mode is more related to the oscillations of the phase of the order parameter than to those of the amplitude of the order parameter.
One can also confirm this claim by observing $\chi_{21}^R(t)$ (not shown).

\subsection{Dynamical phase transition}
\begin{figure}[t]
  \centering
    \hspace{-0.0cm}
    \vspace{0.0cm}
   \includegraphics{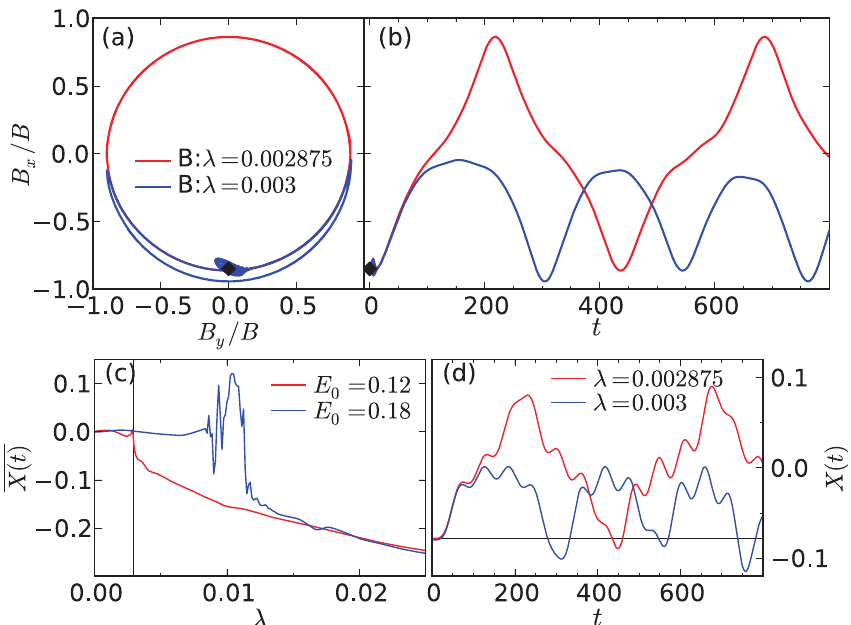} 
  \caption{(a)(b) Trajectory of the pseudo-magnetic field at ${k}=0$ around $\lambda_c$ (the jump) in Fig.~2(d) of the main text at $E_0=0.12$.
  (c) $\lambda$-dependence of the phonon displacement ($X(t)$) averaged over $t\in[0,400]$. (d) Time evolution of $X(t)$ around $\lambda_c$ (the jump) for $E_0=0.12$. The horizontal line indicates the equilibrium value. The parameters of the system and the pump condition are the same as in Fig.~2(d) in the main text.}
  \label{fig:supp_fig3}
\end{figure}

In Fig.~2(d) of the main text, we have shown the $\lambda$-dependence of the  time-averaged 
order parameter ($\overline{|\phi(t)|}$), 
and we observed a sudden change in $\overline{|\phi(t)|}$ at some critical value $\lambda_c$ at $E_0=0.12$. 
Here we show that
this is associated with a qualitative change in the trajectory of the order parameter after the pump, see Figs.~\ref{fig:supp_fig3} (a)(b).
To see this, let us fix the pump strength and change the el-ph coupling.
When the el-ph coupling is small, the U(1) symmetry of the Hamiltonian is weakly broken and the free energy along the phase direction of 
the order parameter is almost flat.
Therefore, the order parameter, as well as $(B^x,B^y)$,
can still rotate, see the result of $\lambda=0.002875$ in Figs.~\ref{fig:supp_fig3} (a)(b).
For stronger el-ph coupling, however,
the potential barrier becomes higher and the order parameter cannot rotate, see the result of $\lambda=0.003$ in Figs.~\ref{fig:supp_fig3}(a)(b).
This change in the trajectory gives rise to
the sudden change of $\overline{|\phi(t)|}$ at $E_0=0.12$, 
which can be regarded as a dynamical phase transition.
In general the trajectory around the transition between the different types of dynamics can be more involved with transient trappings in the potential minima, 
which manifests itself as a spiky structure in the result for $E_0=0.18$ in Fig.~2(d) of the main text.

We note that the dynamical phase transition manifests itself also in other quantities such as the phonon displacement $X$ (see Figs.~\ref{fig:supp_fig3}(c)(d))
and the gap (not shown).
The sudden change in the time average of $X(t)$ is associated with the change of the trajectory as in the case of $\phi(t)$.
When the el-ph coupling is sufficiently weak $X(t)$ oscillates between positive and negative sector.
On the other hand, with stronger el-ph couplings,  $X(t)$ is confined to the sector characterized by the same sign as in the initial state (in the present case negative).

\subsection{Dynamics in the BEC-BCS crossover regime}
 \begin{figure}[t]
  \centering
    \hspace{-0.0cm}
    \vspace{0.0cm}
   \includegraphics{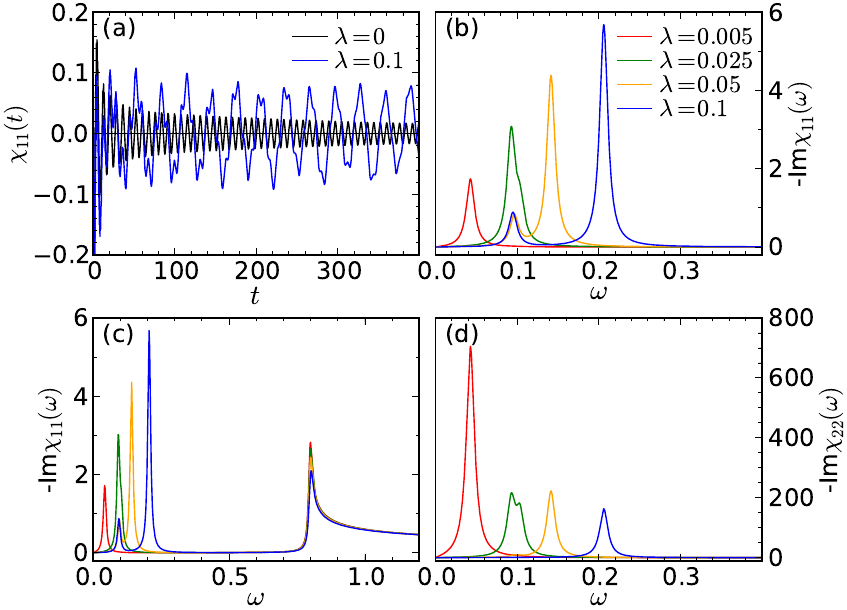} 
  \caption{Susceptibility ($\chi^R_{11}$) in time ((a)) and frequency ((b)(c)) space for various cases with and without phonons for the set 2 parameters.
  (d) Imaginary part of the susceptibility for the phase direction of the excitonic order parameter ($-\text{Im}\chi^R_{22}(\omega)$) for the same condition.
The frequency-dependent quantities $-{\rm Im} \chi^R(\omega)$ are obtained from a Fourier transformation of $\chi^R(t)$ with a damping term $e^{-\eta t}$ and $\eta=0.006$.
  }
  \label{fig:supp_fig3_5}
\end{figure}

In the previous experimental literature on ${\rm  Ta_2 Ni Se_5}$ \cite{Ohta2014}, different model parameters have been extracted from a different ARPES measurement.
Here we show that also for parameters different from those used in the main text the proposed mechanism for the photo-enhanced condensate remains valid.

In the main text, we used the parameters extracted from Ref.~\onlinecite{Mor2016}. 
For this parameter set
the system is on 
the BEC side within the mean-field theory, 
since $B^z_k$ is positive in the EI phase for all $k$ .
On the other hand in Ref.~\onlinecite{Ohta2014}, the ARPES 
spectra show an upturn of the valence band in the EI.
In order to explain this, parameters closer to the BCS-BEC crossover 
have been considered.
The parameters from Ref.~\onlinecite{Ohta2014} are $\Delta_0=0.05,\Delta_1=-2.15,U=2.1$, and $\lambda=0$.
We will use these values, which we will refer to as ``set 2 parameters", as reference parameters below, and adjust the $\Delta_{0,1}$ and $U$ for $\lambda>0$ as explained in the main text.
We consider $T=0$ and $\omega_0=0.1$ and repeat the same analysis as in the main text.
For this parameter set and within the mean-field analysis, $B^z_k$ is negative around $k=0$ in the EI phase, and positive elsewhere.
In this sense, we may say that this is in the BCS regime at least at $T=0$.
The negative $B^z_k$ around $k=0$ leads to the upturn structure in the quasiparticle spectrum, 
hence the minimum band gap $\Delta_{\rm EI}$ is now located away from $k=0$.
This is consistent with the upturn in the ARPES spectrum.
We note that if the difference in the band occupancy $|\Delta n|$ is $1$, which is its maximum value, the Hartree shift is $-U\Delta n=U$, and $B^z_k$  is positive everywhere
for the set 2 parameters. However in the EI phase $|\Delta n|<1$ and this does not happen.

 \begin{figure}[t]
  \centering
    \hspace{-0.6cm}
    \vspace{0.0cm}
   \includegraphics{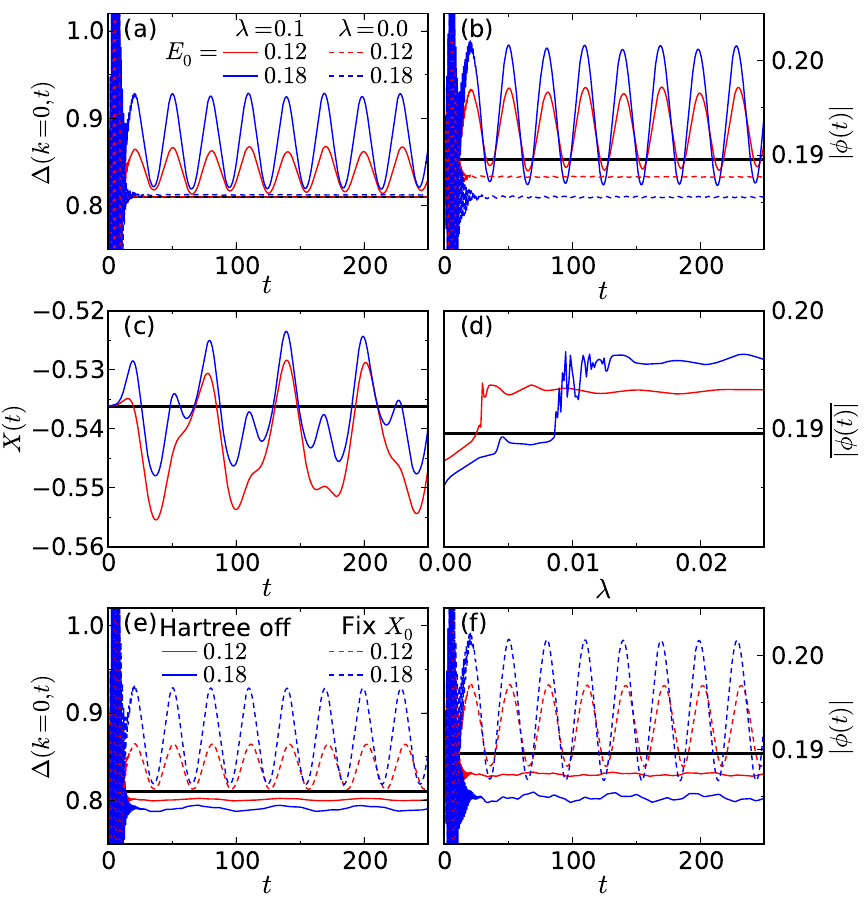} 
  \caption{(a-c)
   Time evolution of the gap at ${k}=0$ ($\Delta(k=0,t)$),
   the excitonic order parameter ($|\phi|$), and the phonon displacement ($X$) for various field strengths and el-ph couplings. (d) The dependence of the size of the order parameter averaged over $t\in[0,400]$ for various field strengths.
  (e,f)  $\Delta(k=0,t)$, $|\phi|$ for $\lambda=0.1$ evaluated by freezing the Hartree shift (solid lines) and the phonon displacement (dashed lines). We have used the set 2 parameters as a reference (see the text).}
  \label{fig:supp_fig4}
\end{figure}
 \begin{figure}[t]
  \centering
    \hspace{-0.6cm}
    \vspace{0.0cm}
   \includegraphics{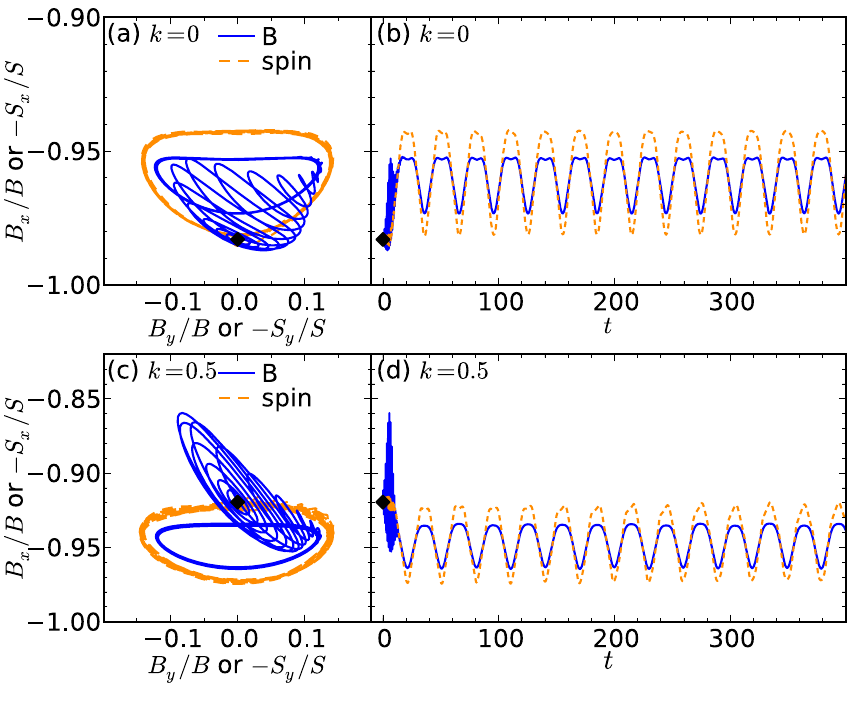} 
  \caption{(a) The trajectory of ($B^x/B,B^y/B$) and ($-S^x/S,-S^y/S$) (b) and the time evolution of $B^x/B$ and $-S^x/S$ at $k=0$ for $\lambda=0.1$.
  (c) The trajectory of ($B^x/B,B^y/B$) and ($-S^x/S,-S^y/S$) (d) and the time evolution of $B^x/B$ and $-S^x/S$ at $k=0.5$ for $\lambda=0.1$.
  We used the set 2 parameters as a reference condition,
  and the parameters $t_p=6.0$ and $E_0=0.12$ for the laser excitation.
  }
  \label{fig:supp_fig5}
\end{figure}
 In Fig.~\ref{fig:supp_fig3_5}, we show $\chi^R_{11}(t)$ for the model with and without phonons. 
Without phonons, there emerge prominent oscillations with the frequency of $\Delta_{\rm EI}=0.80$.
 In contrast to the result in the main text, the damping of the oscillations is well described by 
 a power law 
 $1/t^{0.5}$, which is consistent with the mean-field prediction 
 for superconductors in the BCS regime
  \cite{Volkov1974,Yuzbashyan2006a,Barankov2006,Yuzbashyan2006}.
 This fact shows that
 whether $B^z_k$ changes sign along $k$ in the EI phase has a crucial effect on 
 the decay of the amplitude mode.
 In Figs.~\ref{fig:supp_fig3_5}(b-d), we show the imaginary part 
of the susceptibilities $-\text{Im}\chi^R_{11}(\omega)$ and $-\text{Im}\chi^R_{22}(\omega)$.
 With the el-ph coupling, as in the case in the main text, 
 two additional types of collective oscillations emerge, which originate from the massive phase mode and the phonon.
 The general features of these two modes are 
 the same as in the main text.
 The main difference 
in $-\text{Im}\chi^R_{11}(\omega)$ is the peak and the continuum above $\omega=\Delta_{\rm EI}$, 
 which appears because the amplitude mode with frequency $\Delta_{\rm EI}$  is now prominent and decays slowly.
 
 We further note that if the system 
would be on the verge of the BCS-BEC crossover, one may be able to 
either reveal or suppress the amplitude mode with the frequency $\Delta_{\rm EI}$ by slightly changing the system by, for example, applying pressure or chemical intercalation. 

 \begin{figure}[ht]
  \centering
    \vspace{0.0cm}
   \includegraphics{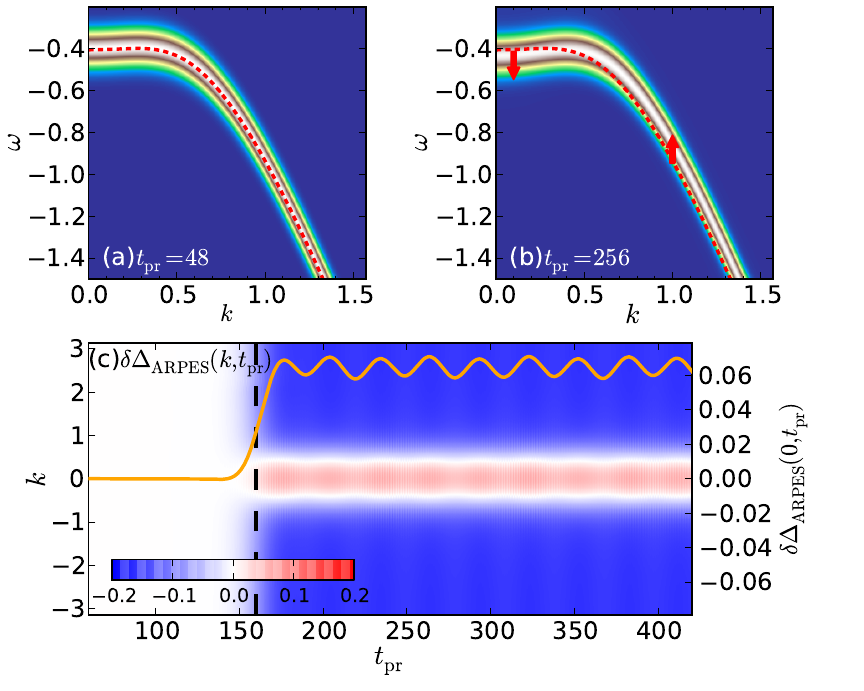} 
  \caption{
  (a)(b) 
  Time-resolved 
  ARPES spectra derived from the retarded part of the Green's functions ($A^R_{k}(\omega;t_{\rm pr})$) before (a) and after (b) the laser pump. Red dashed lines indicate the equilibrium quasiparticle dispersion from the mean-field theory. (c) 
  Time evolution  of the difference between the equilibrium and nonequilibrium gap size at each momentum $k$ ($\delta\Delta_{\rm ARPES}(k,t_{\rm pr})$), see the text. 
  The orange solid line is $\delta\Delta_{\rm ARPES}(0,t_{\rm pr})$ (axis  on the right). 
The parameters are $\lambda=0.1$ with $E_0=0.18$, $\sigma_{\rm p}=3.0$, $\Omega=6.0$, $t_{\rm p}=160.0$, and $\sigma_{\rm pr}=12.0$. 
  The black dashed line indicates the center of the pump pulse.}
  \label{fig:supp_fig6}
\end{figure}
Now we look at the nonequilibrium dynamics after a pump pulse.
Here we use the same condition for the pump as in the main text.
In Fig.~\ref{fig:supp_fig4}, we show the results for the set 2 parameters, which can be compared to Fig.~2
 in the main text.
With the el-ph coupling, we can again see an enhancement of the EI order ($|\phi|$), the displacement of the phonons ($X$) and the gap at $k=0$ ($\Delta(k=0,t)$).
We also confirm that without the Hartree shift there is no enhancement, see Fig.~\ref{fig:supp_fig4}(e)(f).
A positive feedback from the dynamics of phonons again 
exists  for small $E_0$ (compare the full case and the case with $X$ fixed), but it is not as prominent as in the case discussed in the main text.

Without the el-ph coupling, $|\phi|$ decreases after the pump, while $\Delta(k=0,t)$ 
remains almost at the same position.
This originates from  $B^z_{k=0}$ being negative:
After the photo-doping $B^z_{k=0}$ becomes even more negative due to the modified Hartree shift.
Since the gap corresponds to the magnitude of the pseudo-magnetic field (see the explanation around Eq.~(\ref{dwbicbds})), this enhancement of $|B_z|$ and the decrease of the order parameter (which is reflected in $B_{x,y}$) have opposite effects on the size of the gap and compensate each other.

Next we show how the pseudo-magnetic field and the pseudo-spin evolve in the present case.
First we note that $B^z_{k=0}$ is negative and it becomes more negative after the pump. Therefore, the magnetic field is less tilted along the $xy$ direction and we can expect 
a decrease of $|S^x_{k}+iS^y_{k}|$.
This is indeed the case as is depicted in Fig.~\ref{fig:supp_fig5}(a)(b).
On the other hand, away from $k=0$, $B^z$ becomes positive in equilibrium 
and the mechanism mentioned in the main text is applicable again.
As is shown in Fig.~\ref{fig:supp_fig5}(c)(d), one can see 
that there is indeed an 
enhancement of $|S^x_{k}+iS^y_{k}|$ away from $k=0$.
For the set 2 parameters, the region around $k=0$ and that  away from $k=0$ therefore give  negative and positive contributions to the EI order after the pump, respectively, but in total the positive contribution dominates and the EI order is enhanced.
Hence, the mechanism discussed in the main text also holds for the present choice of parameters.

Finally, we show the trARPES spectrum in Fig.~\ref{fig:supp_fig6}. 
This corresponds to Fig.~4 of the main text.
Before the excitation, we can see the slight upturn in the dispersion, see Fig.~\ref{fig:supp_fig6}(a).
This originates from $B^z_k$ being negative around $k=0$. 
After the pump pulse, the band shifts away from the Fermi level around $k=0$, while it shifts toward the Fermi level away from $k=0$.
As a result the upturn becomes more prominent, see Fig.~\ref{fig:supp_fig6}(b).
In Fig.~\ref{fig:supp_fig6}(c), we show $\delta\Delta_{\rm ARPES}(k,t_{\rm pr})$, the time evolution of the difference between the equilibrium and nonequilibrium band gap at each ${k}$.
One can again see the decrease of the band distance around $k=0$, and the increase away from $k=0$.
As can be seen from $\delta\Delta_{\rm ARPES}(0,t_{\rm pr})$, the band position oscillates with the frequency of the collective modes.

\end{document}